\documentclass[%
twocolumn,showpacs,amsmath,amssymb
]{revtex4}

\usepackage{xcolor}
\usepackage{graphicx}
\usepackage{dcolumn}
\usepackage{bm}
\usepackage{cancel}

\begin{document}
	
	
	\title{Compressing a Cylindrical Shell with a Rigid Core }
	\author{Hung-Chieh Fan Chiang, Hsin-Huei Li, and Tzay-Ming Hong$^{\dagger}$}
	\affiliation{Department of Physics, National Tsing Hua University, Hsinchu 30013, Taiwan, Republic of China}

	\date{\today}
	
	\begin{abstract}
	Compressed cylindrical shells are common in our daily life, such as the diamond shape in rolled-up sleeves, crumpled aluminum cans, and retreated package of now defunct drinking straws. The kind of deformation is formally called the Yoshimura pattern. However, there are many other equally prevalent modes of deformation, depending on the relative size of radius between the shell and its inner core, the thickness and rigidity and plasticity of the shell, etc. To elucidate the phase diagram for these modes, we combine molecular dynamics simulations and experiments to study the energetic, mechanical, and morphological responses of a compressed cylindrical shell with a hard core.

	\end{abstract}
	\maketitle
	
	\section{introduction}
\begin{figure}
		\centering
		\includegraphics[width=8.5cm]{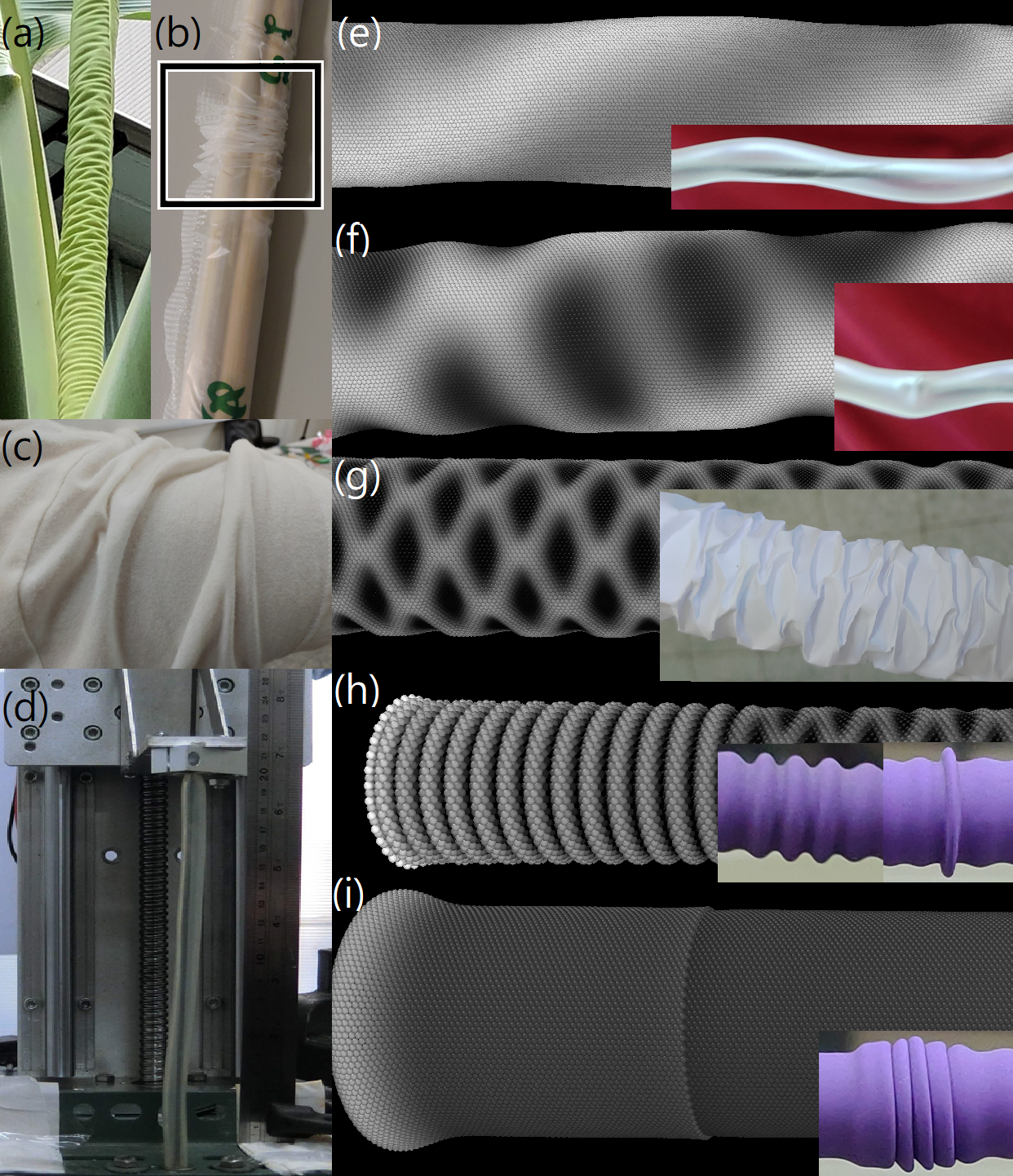}
		\caption{(color online) Examples of compressed cylindrical shell, (a) banana leaf, (b) package of chopsticks, and (c) rolled-up sleeves. Panel (d) shows the experimental set-up. Different modes of compressed shell can be observed: (e) spiral, (f) ladder, (g) diamond, (h) wrinkle, and (i) sagging. The inset of (e, f) are silica gel tube, (g) is paper roll, and (h, i) are balloon. Further detail of inset (h): left part shows wrinkle, and the right part shows ridge.\cite{balloon} }
	\label{mode}
	\end{figure}

	\color{black}
	The deformation of soft material is general and can be due to mechanical force, temperature, pH value, humidity, electric field, and van der Waals interactions\cite{1,2,3,4,5,6,7,8,9,10,11,12}. Plentiful examples exist in living creatures, such as the wrinkling of skins\cite{skin,finger1,finger2}, differential growth of bacterial biofilms\cite{bacterial}, and pattern selection in growing tubular tissues\cite{tubular}. Researchers who were interested at distinct undulating topologies for fruits and vegetables\cite{fruit} or soft elastic cylindrical shell\cite{fold,balloon} have discussed different modes of deformation for core-shell structures. But, in real-life examples of rolled-up sleeves, taken-off pants, shedding skin of snake, (now extinct) wrinkled wrapping when removed from drinking straws, or more exotic banana leaves as shown in Fig. \ref{mode}(a-c), there is a gap between the core and shell. Intuitively this interval lends more freedom to the deformation and may allow for the creation of new modes. To clarify this conjecture we will discuss the morphology and dynamics of a cylindrical shell with a coaxial core whose radii $R_0$ and $R_{\rm in}$ are independently varied. Note that  $R_0$ is defined as the average of inner and outer radii of shell. In addition, the compression rate $v$ and shell properties, such as initial length $L_0$, thickness $t$, and hardness, will be among our tuning parameters.

	We use a stepping motor to compress cylindrical shells of silica gel and paper with a rigid steel rod at its core, as shown in Fig. \ref{mode}(d). To avoid friction, we lubricate the core with a thin layer of oil before each round of compression. In order to better understand the distribution and ratio of bending and stretching energies, molecular dynamics (MD) simulation will also be employed. In contrast to just one mode (diamond) of deformation for stand-alone shells\cite{1977} and two (wrinkle and sagging) for core-shell without a gap\cite{fold, balloon}, our core-gap-shell system can exhibit as many as five different modes. Among them, the spiral, ladder, diamond, and sagging modes show up in thick shells, as shown in Fig.\ref{mode}(e, f, g, i). Whiles,  only  diamond, wrinkle, and sagging modes are observed in Fig.\ref{mode}(g, h, i) for thin shells.

	\section{MD simulation}
	Our MD simulation adopts the Weeks-Chandler-Anderson potential \cite{WCA} to enforce the excluded volume for each lattice point and define the length unit $\sigma$ and the energy unit $\epsilon$. We choose a hexagonal lattice with mean spacing $a_0=1.0\  \sigma$ to form a cylindrical shell. Two impenetrable walls are arranged at both ends of the shell. The reduction of length to  $L$ is carried out by moving one of the walls. Since $k_b/k_s = 3 t^2 /32$ where $k_b$ and  $k_s$ are the bending and stretching modulii \cite{change}, and two values of $t$, $0.23\sigma$ and $2.3\sigma$, are tested by varying $k_b=10^3 \sim 10^5 \epsilon/\sigma^2$ while holding $k_s$ fixed. In the mean time, different materials are simulated by changing $k_s$ for the same $t$. The elastic energy comprises two forms: stretching energy $E_s=k_s(a-a_0)^2/2 $ and bending energy $E_b=k_b(\theta-\theta_0)^2/2 $ where $a$ is the length between adjacent beads and $\theta$ is the angle spanned by three consecutive beads along a lattice direction and $\theta_0=\pi$ along the length and  $\theta_0=[(2\pi R-2)/(2\pi R)]\pi$ in the circular cross section.	Plasticity is included by halving the magnitude of $k_b$  beyond a yield angle $|\theta -\theta_0 |$ of $10^{\circ}$ \cite{plasticity}. 
	The compression rate is set at $v=-10^{-4} \sigma/ v_s$ where $v_s=\sigma (m/ \epsilon)^{1/2}$ is the time unit and $m$ is the mass of bead. All simulations are performed using LAMMPS version 16Mar18\cite{LAMMPS}.

	\section{mode map}
	Since we already know that there are five modes, we can assemble a mode map to characterize their corresponding stages during compression. Inputs from experiments and MD simulation tell us that plasticity promotes diamonds while suppressing other modes, which observation is consistent with our experience with a casually rolled-up sleeve. Readers may protest that sleeves can also assume the neat and more stable mode of sagging. But this is created by deliberate folding and not in the scope of our discussion. The deformation mode of silica gel tube, representative of an elastic shell, is more diversified than its plastic counterpart. The mode map consists of three dimensionless parameters: reduced length $L/L_0$, $R_0-R_{in}/t$, and  $ks/v^2 \times \rho$ that incorporates the effect of hardness and compression rate where $\rho$ denotes the surface mass density of the shell in simulation.

\begin{figure}
		\centering
		\includegraphics[width=8.5cm]{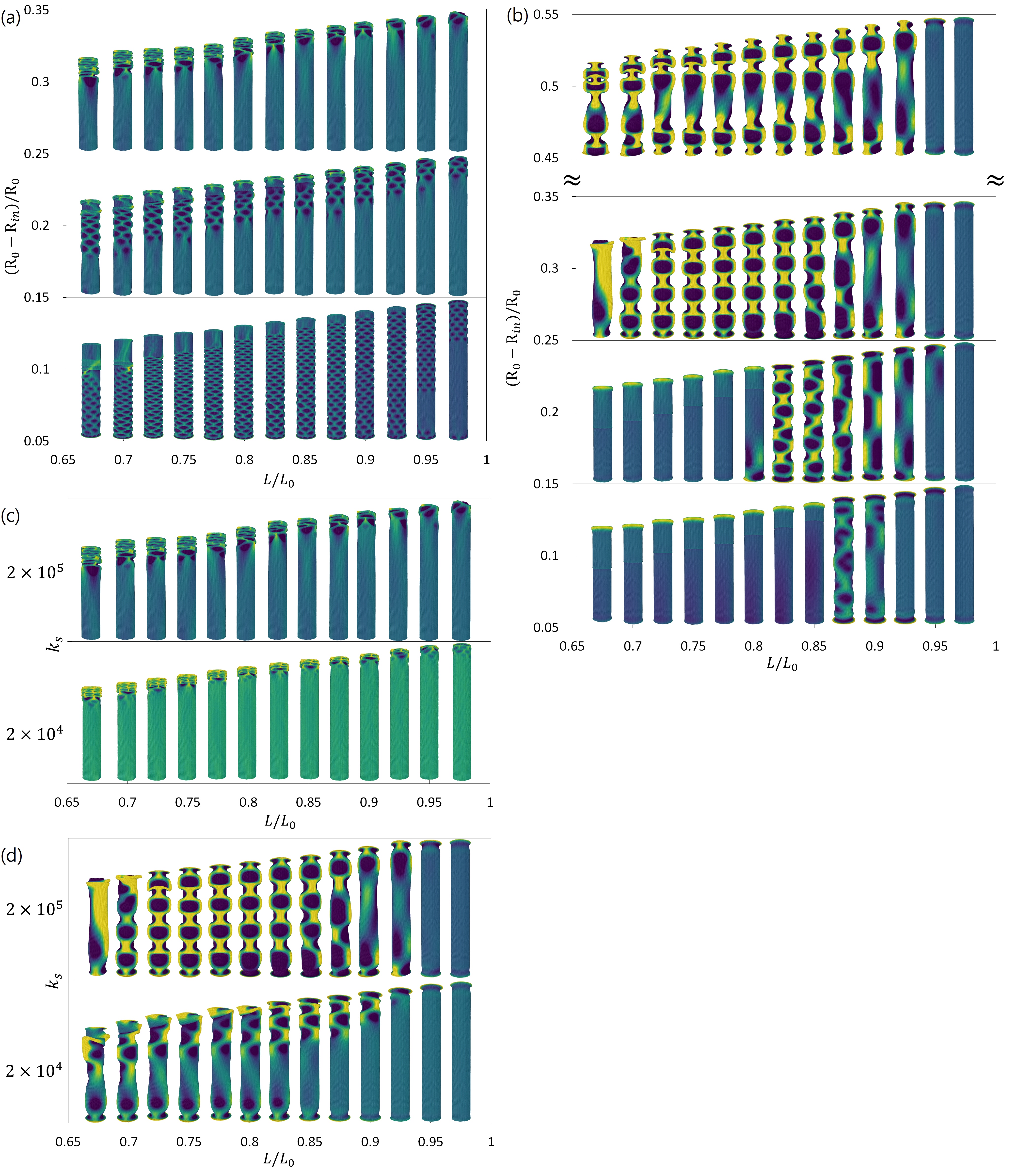}
		\caption{(color online) For panel (a)/(b) shows the deformation mode of shell with $R_0=20$, $L_0=200$, $v=-10^{-4}$, $k_s=2 \times 10^5$, and $t=0.23/2.3$. Different hardness in (c)/(d) with $R_0=20$, $R_{\rm in}=16$ $L_0=200$, $v=-10^{-4}$, $t=0.23/2.3$.}
	\label{map}
	\end{figure}

	\color{black}
	Note that all deformations originate from the moving end. The first mode that appears is spiral that consists of two parallel lines spiraling up the shell
	Further compression introduces two sets of rungs that wrap around the shell and  use the spirals as side rails to form the ladder mode. The rungs appear crooked at first, but improve in their orientation, namely, becoming more parallel as $L$ decreases.  When the gap $R_{0}-R_{\rm in}$ is increased from 2 to beyond 5, a fourth mode of diamond will interpose itself between and coexist with ladder and sagging. But if the gap is further widened to 10, there is no sagging after diamond. A complete survey of modes can be found in Fig. \ref{map} which plots gap vs. $L/L_{0}$. Note that once the sagging shows up, the preceding mode, either ladders or diamonds, will be quickly suppressed to resume the surface back to smoothness. In contrast, the transition from ladder to diamond is gradual.
	
	For an elastic thin shell, diamond is the first mode that emerges. (a) After the diamond pattern gradually spreads and covers the whole shell, wrinkles start to appear and coexist with the diamond. (b) A second transition happens when all the wrinkles are wiped out by a sudden appearance of sagging. (c) Afterwards, wrinkles reemerge between and coexist with sagging and diamond. Further compression will introduce a second sagging and the processes (b) and (c) are duplicated. As shown in Fig. \ref{map}, the number of repetitions can be as frequent as four. We have checked the effect of plasticity which turns out to suppress all the deformation mode except diamond. Furthermore, plasticity renders the arrangement of diamond less periodic.

	\begin{figure}
		\centering
		\includegraphics[width=8.5cm]{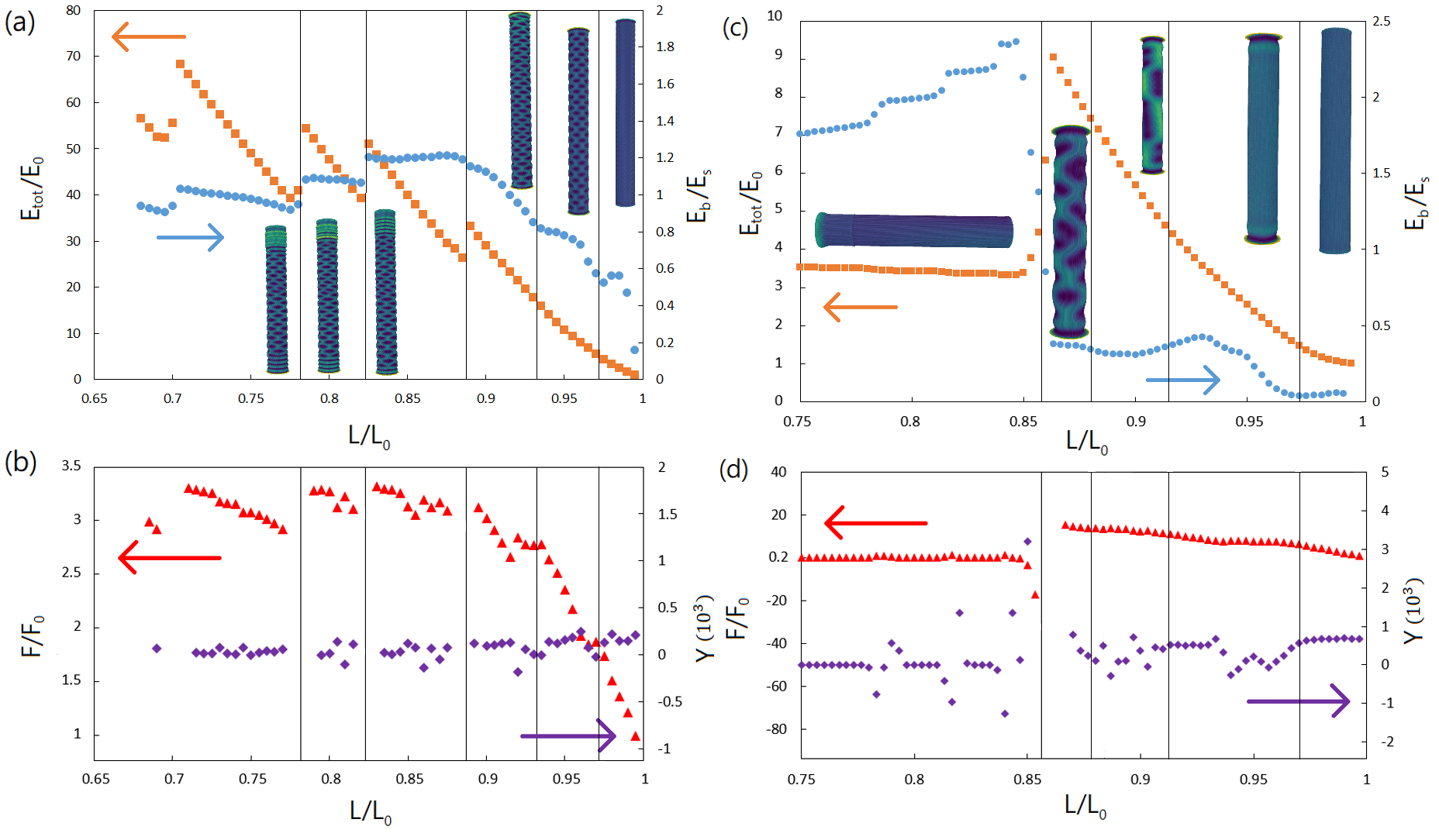}
		\caption{(color online) Panel (a)/(b) shows the energetic/mechanical response of $R_0=10$, $R_{\rm in}=9$, $L_0=200$, $v=-10^{-4}$, $k_s=2 \times 10^5$, $t=0.23$. Each thin straight line highlight the pattern transition on the shell or the formation of the sagging. During the uniform diamond mode, $Y$ and $E_b/E_s$ remain a constant. For panel (c)/(d) shows the the  energetic/mechanical response of $R_0=20$, $R_{\rm in}=18$, $L_0=300$, $v=-10^{-4}$, $k_s=2 \times 10^5$, $t=2.3$. The transition is simpler than thinner shell, the first mode is spiral then turn into ladder and end with sagging. A constant $E_b/E_s$ during ladder mode, and decrease during sagging mode. Besides ladder, spiral and sagging remain a constant $Y$. }
	\label{response}
	\end{figure}
	
	\section{Energetic response}
	The energetic and mechanical responses for a thin shell from MD simulations  are shown in Fig. \ref{response}(a, b) where $E_0$ is the initial total energy and $F_0$ is the resistance force at $L/L_0=0.99$. Being the first mode to appear, diamond increases its size with compression. It was mentioned in Fig. \ref{map} that there are only three modes for a thin shell. Wrinkling emerges and coexists with the diamonds at the first transition, which is characterized by a discontinuous $d(E_b/E_s)/dL$. In contrast, each of the latter transitions introduces one more sagging and exhibits discontinuities in both $E_{\rm tot}$ and $E_b/E_s$. Analogous to the latent heat when water freezes, the difference of $E_{\rm tot}$ is converted into kinetic energy before being dissipated as heat. 
	In contrast to the total $E_b$ and $E_s$ in Fig. \ref{response}(a), we have also measured those for individual mode. The ratio $E_b/E_s$ turns out to be insensitive to $L$ and remains at 3/2 for wrinkles and 0.8/1.4/2.1 for small/medium/large diamonds that appear respectively in $L/L_0 >0.93$, $0.89<L/L_0 <0.93$, and $L/L_0<0.89$. The reason why $E_b/E_s$ increase is that, as the height of diamond perimeter increases with compression, its width narrows to avoid stretching the surface, which implies more $E_b$ is required.
	The ratio  $E_b/E_s$ for sagging is also not a constant. This is expected because $E_b$ is stored mainly on the bending part which remains intact as sagging grows. As a result, the input work is totally converted to $E_s$ on the lengthening overlap part.  

	Figure \ref{response}(b) tells us that the force of resistance is discontinuous at the second and following transitions. The Young's modulus $Y$ can be measured by taking the derivative of $F$ with respect to $L$. 
	The observation that $Y$ equals roughly a constant in Fig. \ref{response}(b) should not be taken too seriously. We believe it is due to the fact that diamond remains the dominant mode for $L/L_0>0.7$. We except that $Y$ will change when sagging take place diamond for further compression.
	
	Now let's study the energetic and mechanical responses for thick shells in Fig. \ref{response} (c, d). 
	As depicted in Fig. \ref{map}, the first deformation to appear is a groove at either end of the shell, that is evidenced by a sudden surge of $E_b$ and $E_s$ with $E_b$ being the dominant energy. The grooves persist after spiral and ladder take turns at occupying the shell. The ratio $E_b/E_s$ for these two modes is determined to be roughly 0.4 and 0.45. Unlike thin shells, sagging does not coexist with other modes, i.e., the thick shell resumes to smoothness as soon as sagging appears, which transition is signified by a discontinuity in $E_{\rm tot}$. As further compression lengthens the sagging, the folded edge where $E_b$ mainly resides remains roughly the same, while $E_s$ increases linearly with $L_0-L$. 
	In the mean time, the shape of folded edge flips between being round and blunt as the edge being pushed by the fictitious wall, which adds an oscillatory part to  $E_b/E_s$ vs. $L/L_{0}$ in Fig. \ref{response}(c).

\begin{figure}
		\centering
		\includegraphics[width=8.5cm]{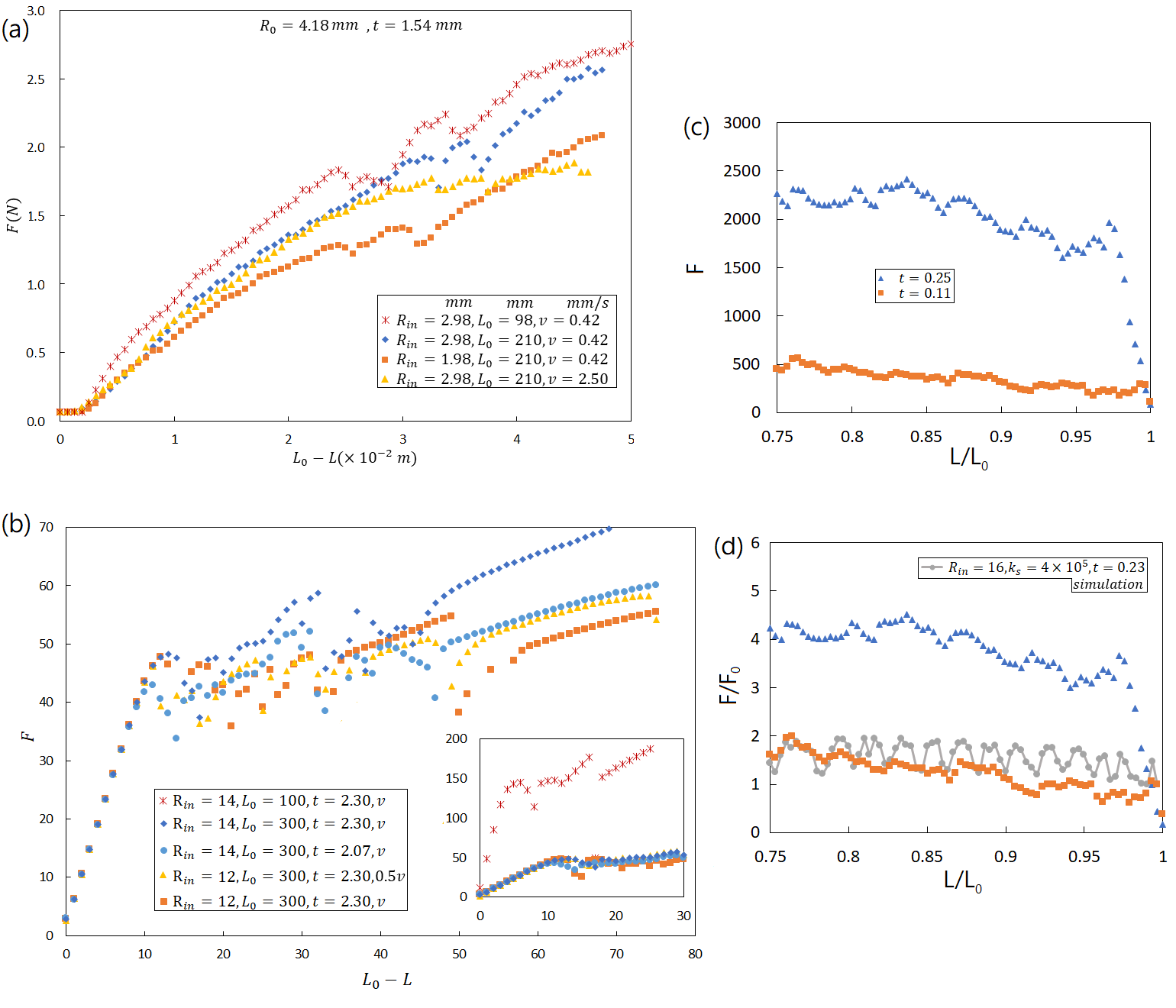}
		\caption{(color online) Resistance force is plotted against reduced length for real samples. (a) Silica gel tube with different $R_{\rm in}$(mm), $L_0$(mm) and $v$(mm/s), while $R_0=4.18$mm is fixed. (b) Simulation result with same variation, while $R_0=20$ is fixed. 
		(b) Paper roll with different $t$(mm) while $R_0=15.9$mm, $R_{in}=11$mm, $L_0=180$mm, and $v$=0.41mm/s are fixed. In contrast to Fig. \ref{response}, plasticity is added to MD simulations with $R_0=20, L_0=300$, and $v=10^{-4}$ to obtain plot (c)  in order to compare with (a) and (b).}
	\label{exp}
	\end{figure}

	Experimental results for silica gel tube are plotted in Fig.\ref{exp}(a) where each dip represents the appearance of a new sagging. A lower compression rate $v$  exhibits a higher $F$. 
	The shorter shell, as denoted by triangles, exhibits a steeper slope than diamonds in  Fig.\ref{exp}(a). This is expected from $F=(YA/L_0 )(L_0-L)$. However, if multiplied by their respective $L_0$, the slope for triangles will become slightly smaller than that for diamonds. This implies that, unlike homogeneous material, $Y$ is not pure an intrinsic property, but may depend on $L_0$ for a shell with deformations.
	The effect of $R_{\rm in}$ is also checked. The fact that the cube data overlap with diamonds before sagging implies that, as long as $R_{\rm in}$ is nonzero, $Y$ is insensitive to its value. And, once entered, sagging always enhances $Y$ besides the induction of a sudden drop in $F$ at the transition.

	Our simulation results without plasticity in Fig.\ref{exp}(b) show that a shorter $L_0$, larger $R_{\rm in}$,  and higher $v$ require a stronger $F$ to achieve the same $L_0 -L$. 	These trends are consistent with those of silica gel in Fig.\ref{exp}(a). 
	One further parameter that we tested in simulations is the effect of thickness $t$. As expected, thicker samples also require a larger force.
	To test the effect of plasticity, we also arrange to sample on paper roll to obtain  Fig.\ref{exp}(c) and contrast it with simulations with plasticity in Fig.\ref{exp}(d). Two characteristics that are unique to plastic shells are that the mode of deformation is reduced to one, i.e., diamonds, and rings of diamonds emerge discretely. The $F$ is found to fluctuate whenever a new ring of diamonds appears, similar to the behavior of paper roll. But our simulations fail to predict the overall increase of $F$ as $L_0-L$ increases. Since there was no such a trend in another experiment\cite{1977} without a core, We believe the increment in $F$ must be due to the increasing number of contacts and frictional force between the shell and the core.

	\section{release}
\begin{figure}
		\centering
		\includegraphics[width=7cm]{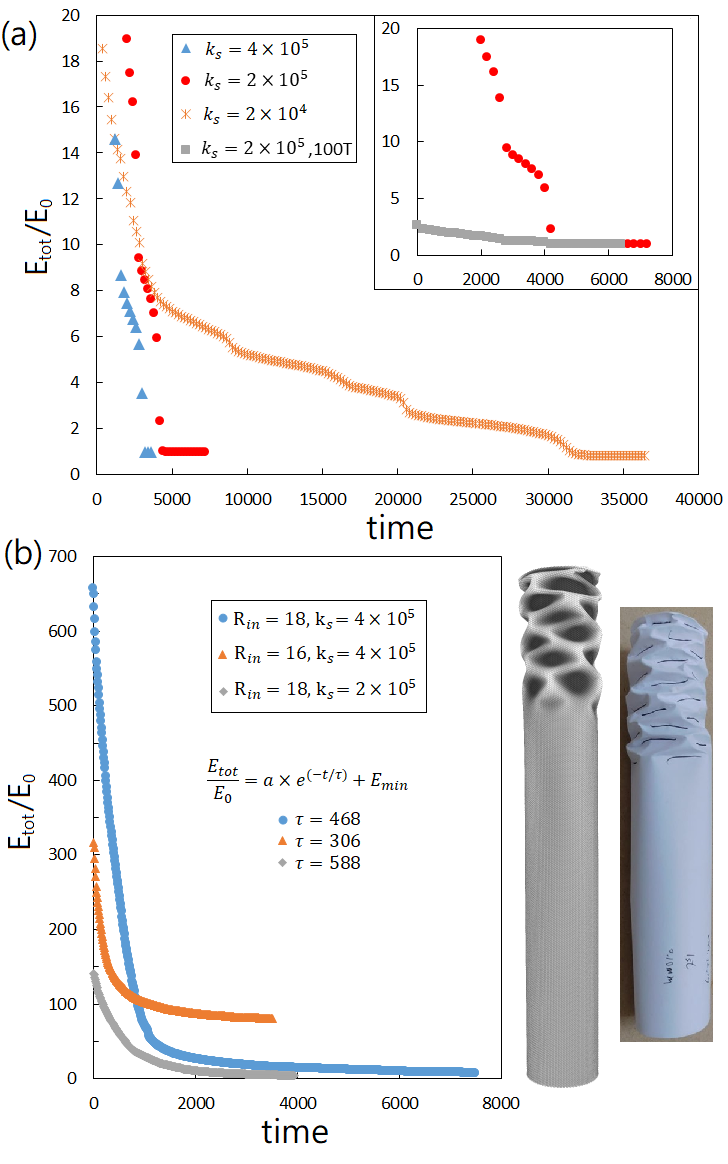}
		\caption{(color online) We shift the time axis to $0$ when the relaxation starts, and distinguish the (a) elastic and (b) plastic case. (a) shows the harder material recover faster with $R_0=20, R_{in}=16, L_0=200, t=0.23$, and independent of temperature in the inset. (b) compares the deformation of simulation and experiment with plastic at the right side, and shows larger space, i.e. $R_0-R_{in}$, recover faster with $R_0=20, L_0=300, v=10^{-4}, t=0.23$. }
	\label{relax}
	\end{figure}

	After we roll up the sleeve, the deformation can persist for a while due to its friction with our arm. How about the compression shell for a purely elastic shell in MD simulations? In other words, is the compression process intrinsically irreversible? Or is plasticity crucial to induce hysteresis?
	This is the goal of our following simulations in which the compressing wall is suddenly removed. The recoil speed $v_{\rm recoil}$ and total energy are recorded and plotted in Fig.\ref{relax}(a). It can be seen that $v_{\rm recoil}$ varies with modes and is the smallest for sagging.  

	The thick shell in MD will always stuck at sagging and never recover to smooth again. We use $E_tot/E_0$ as the degree of deformation to plot the relation to the simulation time $t$ for thin shell in MD. We find that harder and more space can release faster, and independent of temperature. The temperature represent the kinetic energy of the beads, which means the fluctuation of the shell is independent of the relaxation in simulation. To avoid the shell brittle or melt, it can not change the range of temperature to check this property in experiment. For elastic case, as shown in Fig. \ref{relax}(a), the stepped energetic response represents the untying of the sagging. Note that if and only if $R_0-R_{in}>3$, the sagging can be untied.  Compare to the plastic case, as shown in Fig.\ref{relax}(b), it is hard to form sagging during compression, which means that there will no stepped energetic response and there is a good agreement, with $R^2 \sim 0.99$, of the relation: $E_{tot}/E_0 = a \times {\rm exp}(-t/\tau)+E_{min}$, for $a$ is dependent on the maximum total energy of each different case, $\tau$ is relaxation time and $E_{min}$ is $E_{tot}/E_0$ at the equilibrium state.

	\section{discussion}
	Although the morphological and mechanical responses  obtained by our simulation are consisting with experimental results, some discrepancies still exist. For instance, the diamond mode that appears in simulations for a thick shell with  $R_{in}/R_0 >0.6$ was never observed in experiments. In simulations for thick shells the appearance of sagging will wipe out whatever modes preceeding it. However,   in real experiments sagging is less exclusive. 
	
	Boundary condition, e.g., whether the circular ends are allowed to deform during compression, is checked not to alter our conclusions, except minor quantitative corrections to the critical $L/L_0$ when different modes switch. Intuitively one would imagine $R_{in}$ not to matter when the deformation is not serious and the shell never touches the core. However, this is limited to the early stage of compression when $L/L_0 >0.95$. As a result, it is still acceptable for us to use the parameter $(R_0-R_{in})/R_0$ as a label for the mode map.
	
We acknowledge the financial support from MoST in Taiwan under grants 105-2112-M007-008-MY3 and 108-2112-M007-011-MY3.

\end{document}